%% file: paper.tex
\documentclass[lettersize,journal]{IEEEtran}

\input{header.tex}

\begin{document}

\title{CAFÉ, an automated feedback tool to approach Formal Methods}

\author{
  \IEEEauthorblockN{
    G\'eraldine Brieven\orcidlink{0000-0003-1410-1470},Ayman Labrahimi Kasdaoui,Benoit Donnet\orcidlink{0000-0002-0651-3398}
  }

  \IEEEauthorblockA{
    Universit\'e de Li\`ege, Institut Montefiore, Belgium
  }
}

\maketitle

\input{abstract.tex}


\input{intro.tex}
\input{related.tex}
\input{activity.tex}
\input{cafe.tex}
\input{eval.tex}
\input{extension.tex}

\input{conclusion.tex}

\bibliographystyle{IEEEtran}
\bibliography{Bibliography}

\appendix

\input{appendix}

\end{document}

%% file: header.tex
\usepackage{xspace}
\usepackage{nth}
\usepackage{url}
\usepackage{orcidlink}
\usepackage{multirow}
\usepackage{graphicx}					
\usepackage{subfig}
\usepackage{caption}
\usepackage{array} 
\usepackage{amsmath} 
\usepackage{tablefootnote}
\graphicspath{{.}{./Pictures/}}

\newcommand{\dfn}[1]{\textit{#1}}            
\newcommand{\invariant}{loop invariant\xspace}
\newcommand{\invariants}{loop invariants\xspace}
\newcommand{\gli}{\textsc{Gli}\xspace} 
\newcommand{\fli}{\textsc{Fli}\xspace} 
\newcommand{\gliFull}{Graphical Loop Invariant\xspace}

\newcommand{\glibp}{\textsc{Glibp}\xspace}
\newcommand{\glibpFull}{\gliFull Based Programming\xspace}
\newcommand{\bgli}{fill-in-the-blank \gli}

\newcommand{\cafe}{\textsc{Caf\'e}\xspace}
\newcommand{\progchall}{programming challenge\xspace} 
\newcommand{\progchalls}{programming challenges\xspace}
\newcommand{\solPart}{solution part\xspace}
\newcommand{\solParts}{solution parts\xspace}
\newcommand{\SolParts}{Solution parts\xspace}
\newcommand{\var}[1]{\texttt{#1}\xspace}
\newcommand{\initState}{\textsc{Initial State}\xspace}
\newcommand{\inLoopState}{\textsc{In-loop State}\xspace}
\newcommand{\finalState}{\textsc{Final State}\xspace}
\newcommand{\critere}{exit condition\xspace}

\newcommand{\greenBoxes}{\dfn{constrained boxes}\xspace}
\newcommand{\GreenBoxes}{\dfn{Constrained boxes}\xspace}
\newcommand{\RedBoxes}{\dfn{Free boxes}\xspace}
\newcommand{\redBoxes}{\dfn{free boxes}\xspace}

\newcommand{\fliFull}{Formal Loop Invariant\xspace}
\newcommand{\regle}[2]{\textcolor{#1}{#2}\xspace}
\newcommand{\size}{\mbox{\tt N}}

\newcommand{\maxindex}{\mbox{\tt index}\xspace}
\newcommand{\arraya}{\mbox{\tt a}}
\newcommand{\counteri}{\mbox{\tt i}\xspace}

\newcommand{\vLine}[1]{\color{#1}\vrule} 

\usepackage{tikz}
\usetikzlibrary{positioning,decorations.pathreplacing,calc,fit,shapes.geometric, arrows,shapes.symbols,arrows.meta}

\definecolor{MACRouge}{RGB}{255,38,0}
\definecolor{dline}{RGB}{255,38,0}
\definecolor{done}{RGB}{4,51,255}
\definecolor{todo}{RGB}{0,143,0}
\definecolor{lowerbound}{RGB}{255,147,0}
\definecolor{upperbound}{RGB}{255,64,255}
\definecolor{variter}{RGB}{146,144,0}
\definecolor{vardata}{RGB}{169,169,169}
\definecolor{varsize}{RGB}{148,23,81}
\definecolor{unmodified}{RGB}{0,150,255}

\newcommand\tikzmark[1]{%
  \tikz[remember picture] \node (#1) {};
}

%% file: abstract.tex
\begin{abstract}

We present \cafe, a learning platform designed to introduce computer science students to Formal Methods (FM). \cafe aims to scaffold students' \textit{structural} thinking (in contrast with \textit{operational} thinking) by promoting the practice of \glibpFull (\glibp). In the \glibp approach, students solve loop-based problems by first constructing a \gliFull (\gli) before deriving the corresponding code. The \gli is an informal diagrammatic representation of the loop invariant. It illustrates the variables involved in the loop, their properties, and the relationships between them. To enable automated feedback, students complete a \bgli, a box-based version of the \gli. Beyond evaluating the code students submit, \cafe provides personalized feedback on students' \gli and its alignment with the code. In this demo, we walk through \cafe from both a student's and a teacher's perspective. We show how the tool supports \gli design and provides feedback. We then present how a teacher can encode \progchalls together with a corresponding \bgli.

\end{abstract}

%% file: intro.tex
\section{Introduction}\label{intro}

Formal Methods (FM) thinking is a mindset computer scientists should develop from the very first day of their studies~\cite{3_FormalThinking}. Other researchers~\cite{moreFM,moreFMTea,FMimportant} have also highlighted the importance of FM in education, given its increasing adoption in industry and the persistent challenge of writing correct programs.

To address this, we introduced a \glibpFull (\glibp) approach in our introductory programming (CS1) course. It aims to familiarize students with structural thinking early in their education, thereby preparing them for FM. \glibp is grounded in Hoare's logic~\cite{hoare} and the programming methodology established by Dijkstra~\cite{dijkstra}. To make invariants accessible, this approach uses informal visual representations and textual descriptions, referred to as a \gliFull (\gli). A \gli illustrates the objects and variables involved in a loop solution, as well as the relationships between them, and must hold at each evaluation of the loop guard. Students are expected to implement their loops in accordance with their \gli, under the constraint that the loop terminates.

To support regular practice, we developed a learning platform called \cafe. Throughout the semester, students use \cafe to solve loop-based problems by submitting both the \gli reflecting their solution and the corresponding C code. Beyond evaluating the final code, \cafe provides personalized feedback on the \gli and its consistency with the code. This automated feedback guides students in refining their \gli and code before resubmitting for further feedback. In our demo, we use one problem as a running example, showing how it can be solved by completing a \bgli and deriving the corresponding code. We also give the audience hands-on access to \cafe, first from the student's perspective, walking through the automated feedback a student might receive. We then shift to the teacher's view, demonstrating how instructors can encode new \progchalls and define a corresponding \bgli.

\cafe's distinguishing feature is its ability to automatically generate feedback on students' \gli within an interactive online environment.

\subsection{Targeted Audience}

This demo will primarily interest CS educators who teach FM or wish to prepare their students for it. Our tool may also serve as a valuable complementary resource for students who struggle to grasp the role of assertions in program correctness.

More broadly, we are open to extending our automated feedback feature to support the teaching of FM in other course contexts. We have just developed a proof-of-concept prototype in which a \gli and a C code snippet are automatically translated into Dafny for formal verification. We would warmly welcome collaboration with other universities on this project.

%% file: related.tex
\section{Related Work}\label{sec:related}

{\setlength{\doublerulesep}{0pt}
\setlength{\tabcolsep}{5pt}
\begin{table*}[!h]
  \caption{\centering Examples of the \progchalls supported by \cafe during the academic year 2025--2026.}
  \label{tab:challenges_list}
  \begin{center}
    \resizebox{\textwidth}{!}{%
    \begin{tabular}{lll}
      \multicolumn{1}{c}{\textbf{ID}} & \multicolumn{1}{c}{\textbf{Main problem description}} & \multicolumn{1}{c}{\textbf{Subproblems (or modules)}}\\
      \hline\hline\hline\hline
      \multirow{2}{*}{1} &  \multirow{2}{*}{Computing the product of the non-zero digits for all numbers in $[\var{a}, \var{b}]$} & outer loop (iterating over the range $[\var{a}, \var{b}]$)\\
        &  & inner loop (computing the product of non-zero digits in a given number)\\
      \hline
      2 & Compressing an integer array into another & a single loop (compressing the array)\\
      \hline
      \multirow{4}{*}{3} & \multirow{4}{*}{Displaying all numbers in array \var{T} that follow a property $G$} & \texttt{uint nb\_digits(uint x)}\\
        & &  \texttt{uint factorial(uint x)}\\
        & &  \texttt{uint is\_g(uint x)}\\
        & &  \texttt{void display(int T[], int N)}\\
    \end{tabular}
    }
  \end{center}
\end{table*}
}

The foundational work on \invariant-based programming was established by Dijkstra~\cite{dijkstra}, followed by significant contributions~\cite{gries_book,morgan_book}. 
They represent invariants as logical assertions. 

A few years later, several educational approaches have emerged to teach \invariant-based programming. Tam~\cite{teaching_loop_invariant} advocates introducing students to \invariants early in programming courses and provides examples of code construction using informal \invariants expressed in natural language. Astrachan~\cite{invariants_pictures} suggests using graphical \invariants (\gli) in introductory computer science courses (CS1/CS2), though specific states are not represented, contrary to our approach (detailed in Subsection~\ref{sec:glibp}). To our best knowledge, Astrachan is the first researcher to use diagrammatic reasoning~\cite{Anderson2002} to make \invariant more accessible. 
More recently, Mannila~\cite{invariant_edu} developed a \invariant-based programming approach (IBP) that aligns more closely with our \textit{\textbf{GL}IBP} methodology. Their approach targets novice programmers through visual program construction designed to minimize notational overhead. Like our approach, they represent \invariant states visually. The key distinction is that we employ textual notations while they use predicates within diagrams.

To facilitate the adoption of invariant-based programming, some tools~\cite{loop_invariant_misconceptions,tool_proofs} have been developed. They provide scaffolding for students for writing proofs. However, they do not support \textit{graphical} loop invariants and automated feedback on them, like our learning platform \cafe~\cite{cafe2}.

%% file: activity.tex
\section{Context}\label{sec:context}

\subsection{Our CS1 course and the role of \cafe in the course}

Our experience aligns with Morazán's observation~\cite{li_otherApproachWithRec}, who identifies the correct construction of loops as the primary difficulty faced by novice programmers. Therefore, we introduce \glibp early in our CS1 course, at the same time as loops themselves. At that point, students have not yet encountered functions, which is why we do not include pre- and post-conditions in our problem-solving approach. 

Our \glibp methodology spans roughly a third of the course schedule, using C as the programming language.
Additionally, students practice \glibp on \cafe through three \progchalls. This typically requires an additional five hours of out-of-class work. Each \progchall consists of a loop-based problem, potentially broken down into several subproblems (defined in the last column of Table~\ref{tab:challenges_list}). For each subproblem (relying on a loop), a \bgli and a code template are provided (Fig.~\ref{cafe.bgli}). Students have a 2-day window to submit their solution up to three times, each submission receiving automated feedback. The latest submission determines the final mark, and each \progchall accounts for 2\% of the final course grade. After this certificative period, students are free to keep training, but this no longer affects their final grade.

\subsection{\glibpFull (\glibp)} \label{sec:glibp}

\begin{figure}[!h]
  \begin{center}
    \includegraphics[width=0.8\linewidth]{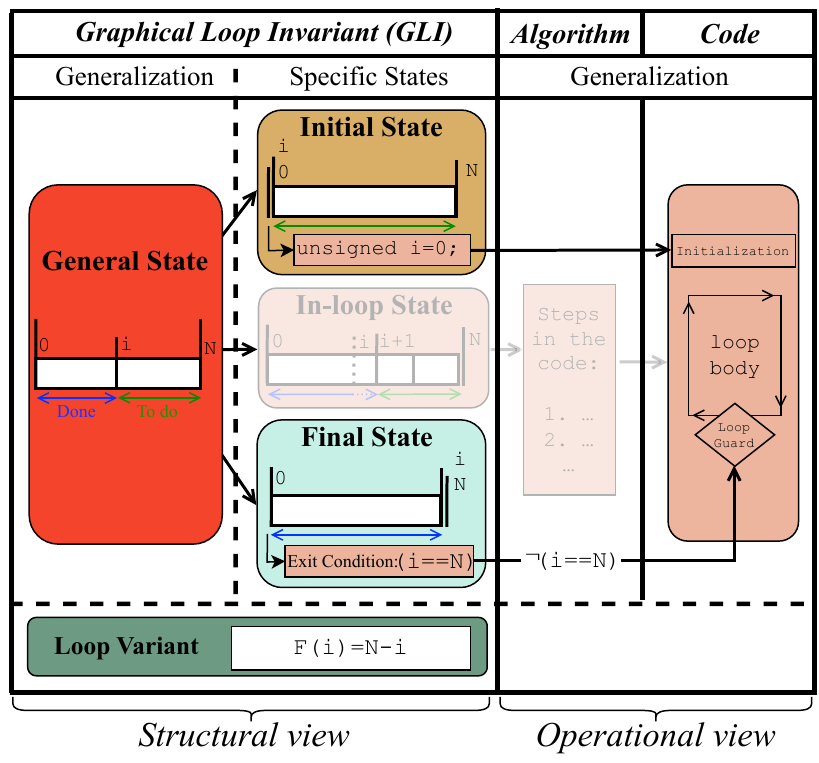}
  \end{center}
  \caption{\centering \SolParts of the \glibp approach. Neither the \inLoopState nor the steps of the program are explicitly required from students, which is why they are blurred.}
  \label{glibp.overview.fig}
\end{figure} 

In this section, we describe the \glibp approach taught in our CS1 course~\cite{4_glibp}.
It applies whenever students need to implement a loop, relying on an informal version of the invariant: the \dfn{\gliFull} (\gli). The \gli must depict the object(s) being iterated over (e.g., an array or a range of integers), the variables involved in the loop, their domains, and the relationships among them that must be maintained across all iterations.
Fig.~\ref{glibp.overview.fig} shows that a \gli includes a general state that can be instantiated into initial and final states. These three diagrams support a structural view~\cite{structuralView} of the solution. Next, as illustrated in Fig.~\ref{glibp.overview.fig} in the ``Code'' column, students derive the variable initialization, the loop guard, and the loop body, shifting to an operational view. Additionally, they must provide a \dfn{loop variant} function to justify that the loop terminates.

\subsection{Formal semantics to provide automated feedback on a Graphical Loop Invariant} \label{sec:notations}

To enable automated feedback, the \gli had to be formalized. To do so, we defined its components, specific boxes to host them, and a set of rules they should meet~\cite{4_glibp}. Among its components, a \gli should include: the name of the structure(s) to iterate over, the lower and upper bounds, the dividing line(s) tagged with an iteration variable, and the description of each zone already browsed by the dividing line(s)\footnote{This expresses that the elements of this zone have already been treated in the loop.}. These descriptions express the property these elements meet and may include accumulators. The rules are parametrized through a configuration file that sets, via the boxes, the expected variable names, relationships between them, etc.

\begin{figure*}[!h]
  \begin{center}
    \includegraphics[width=0.7\linewidth]{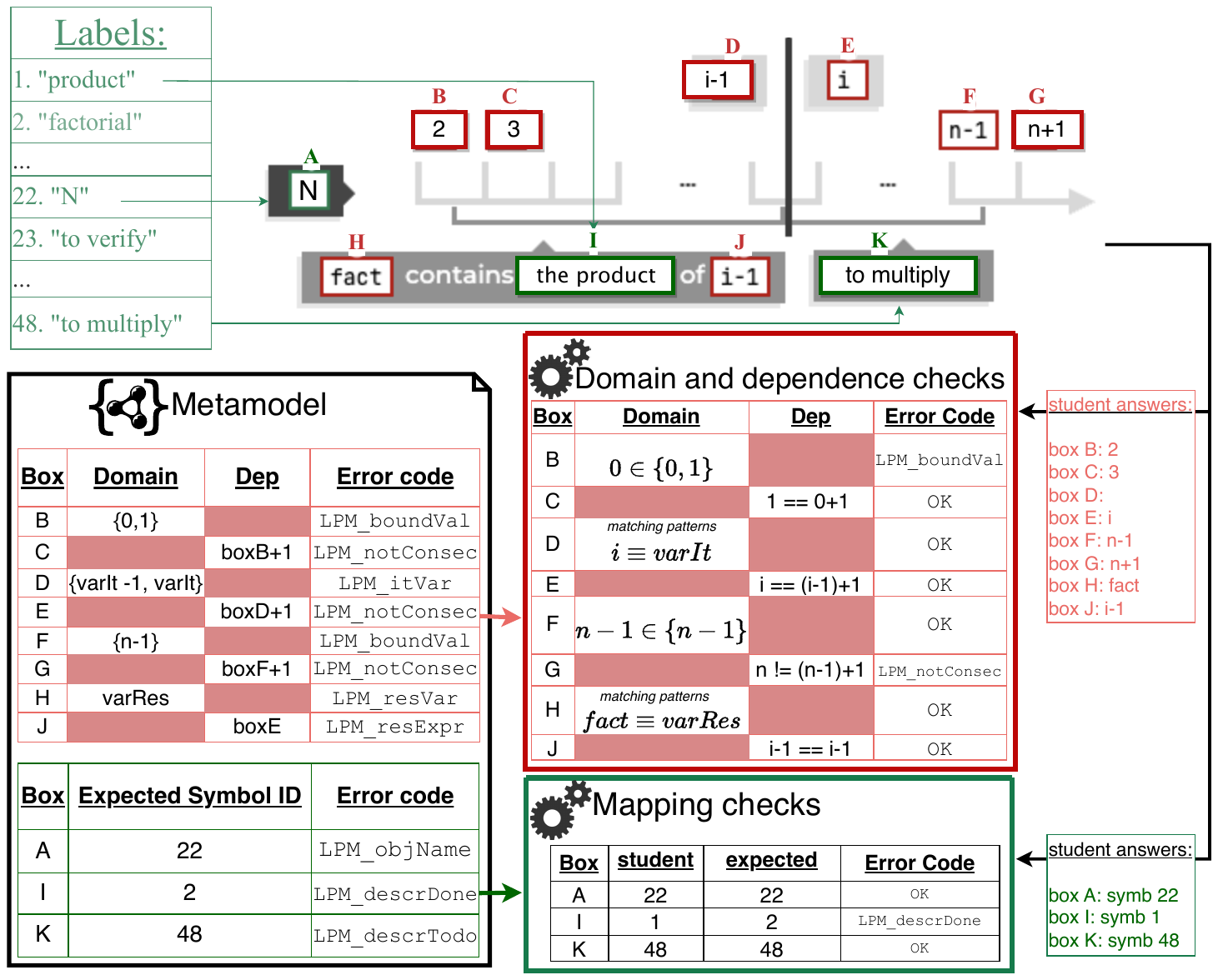}
  \end{center}
  \caption{\centering Example of \bgli, metamodel, and checking rules in CS1. Syntactic checks on \redBoxes are omitted for figure clarity.}
  \label{fbGen.bgli_checker}
\end{figure*}

Fig.~\ref{fbGen.bgli_checker} illustrates a \bgli containing two types of boxes. \RedBoxes (in red) expect expressions, allowing students to define their own variables and apply operations to them. \GreenBoxes (in green) expect a symbol chosen from a predefined list.
In the configuration file, a metamodel is defined, linking each box to a typical error. The typical errors are defined beforehand. For the checks, \redBoxes are mapped to a domain or to a dependency relationship with other boxes, while \greenBoxes are mapped to a symbol $ID$. For instance, box $E$ should contain the expression in box $D$ plus 1, and box $K$ should hold the \nth{48} symbol in the list (i.e., ``to multiply''). In this example, the student made three mistakes, in boxes $B$, $G$, and $I$. In box $B$, the student entered 2, while the expected values were 0 or 1; this triggers the error code \texttt{LPM\_boundVal}. Similarly, in box $G$, the expected content should match the content of box $F$ minus 1 (i.e., \texttt{n}), which is not the case. Lastly, in box $I$, the expected content was the second symbol (``factorial''), but the student selected the first one (``product'').

\subsection{How the \gli Helps Students Learn to Write Formal Specifications}

\begin{figure*}[h!]
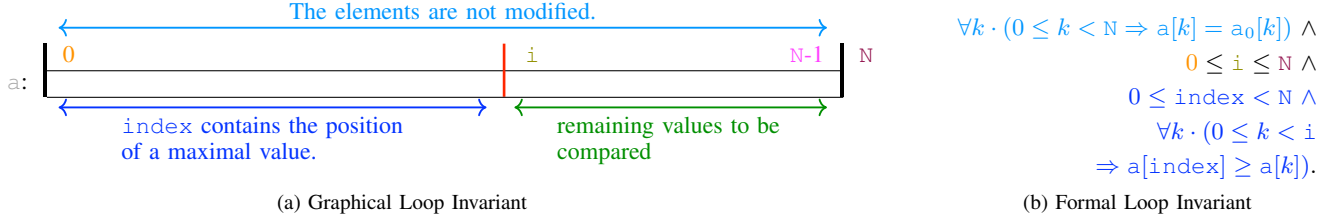

  \centering
  \subfloat[\centering \gliFull]{%
    \label{fig:GLI_findMax}
    \begin{minipage}[t]{0.6\textwidth}
      \vspace{0pt}
      \centering
      {\small
      \begin{tabular}{lm{2cm}crm{1.6cm}crm{2cm}crl}
        & & & & & & & & & & \\
        & \tikzmark{a} & & & & & & & & \tikzmark{b} & \\
        \multicolumn{1}{{r!{\vLine{black} width 1.2pt}}}{} & \multicolumn{1}{l}{\regle{lowerbound}{0}} & & & &  & \multicolumn{1}{{!{\vLine{MACRouge} width 1pt}c}}{\regle{variter}{\var{\counteri}}} & & & \multicolumn{1}{{r!{\vLine{black} width 1.2pt}}}{\regle{upperbound}{\var{\size}-1}} & \regle{varsize}{\size}\\
        \cline{2-10}
        \multicolumn{1}{{r!{\vLine{black} width 1.2pt}}}{\regle{vardata}{\arraya}:}  & & & & & &  \multicolumn{1}{{!{\vLine{MACRouge} width 1pt}c}}{} &  &  & \multicolumn{1}{{r!{\vLine{black} width 1.2pt}}}{}  &   \\
        \cline{2-10}
        & \tikzmark{c} & & &   & \tikzmark{d} & \tikzmark{e} &  &  & \tikzmark{f} & \\
      \end{tabular}

      \tikz[remember picture,overlay]
        \draw[<->, line width=0.25mm, unmodified]
          ([xshift=-0.05cm]a.west) --
          node[above]{\regle{unmodified}{The elements are not modified.}} 
          ([xshift=0.1cm]b.east);

      \tikz[remember picture,overlay]
        \draw[<->, line width=0.25mm, done, text width=4cm]
          ([xshift=-0.05cm]c.west) --
          node[below, align=left]{\regle{done}{\maxindex contains the position of a maximal value.}}
          ([xshift=0.1cm]d.east);

      \tikz[remember picture,overlay]
        \draw[<->, line width=0.25mm, todo, text width=3cm]
          ([xshift=-0.05cm]e.west) --
          node[below]{\regle{todo}{remaining values to be compared}}
          ([xshift=0.1cm]f.east);
      }
    \end{minipage}
  }
  \hfill
  \subfloat[\fliFull]{%
    \label{fig:FLI_findMax}
    \begin{minipage}[t]{0.3\textwidth}
      \vspace{0pt}
      \centering
      {\small
      \begin{align*}
        \regle{unmodified}{\forall k \cdot (0 \leq k < \size \Rightarrow \arraya[k] = \arraya_0[k])}\ \wedge \\
        \regle{lowerbound}{0} \leq \regle{variter}{\counteri} \leq \regle{varsize}{\size}\ \wedge \\
        \regle{done}{0 \leq \var{\maxindex} < \var{\size}\ \wedge}   \\
        \regle{done}{\forall k \cdot (0 \leq k < \counteri} \\ \regle{done}{\Rightarrow \arraya[\maxindex] \geq \arraya[k])} .
      \end{align*}
      }
    \end{minipage}
  }
  \caption{\centering Loop invariant for finding a maximal value in an array.}
  \label{fig:findMaxLoopInvariant}
\end{figure*}

In the context of the CS2 course, when students are introduced to formal specifications, they derive them from a \gli. They translate each property carried by the diagram, optionally using an intermediate translation table, as illustrated in Table~\ref{tab:translation}. This table serves as an intermediary between the two representations of the loop invariant given in Figure~\ref{fig:findMaxLoopInvariant}.

\begin{figure}[!h]
  \begin{center}
    \includegraphics[width=0.9\columnwidth]{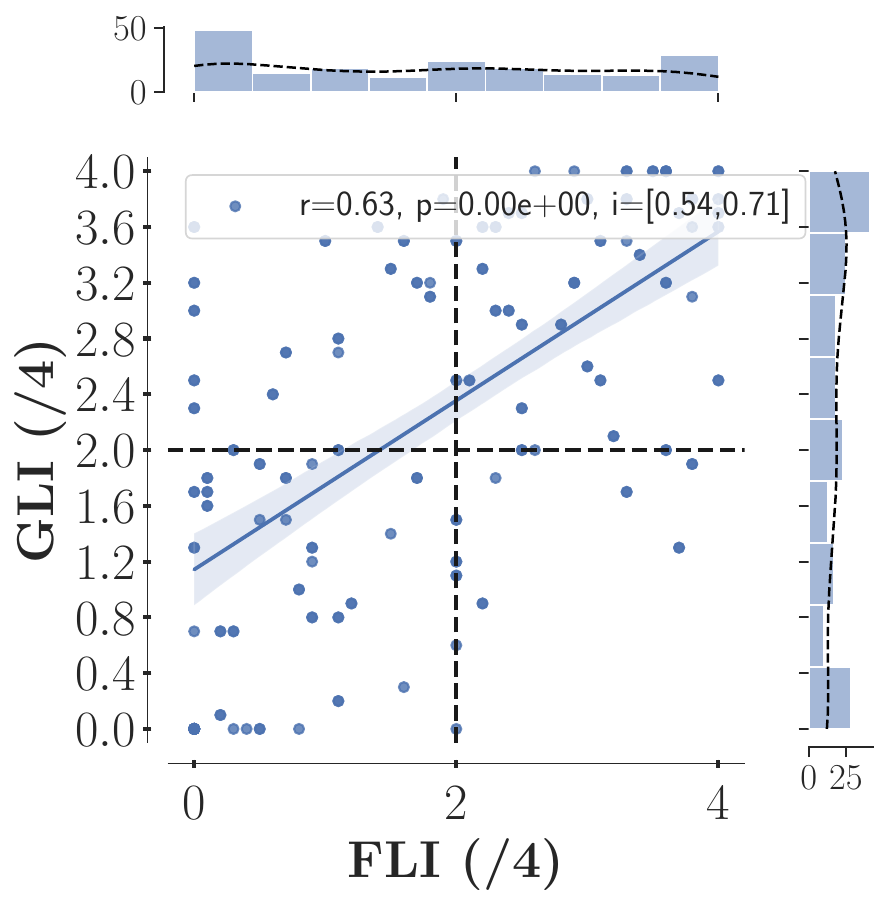}
  \end{center}
  \caption{\centering Correlation between students' grades in the \gli and the Formal Loop Invariant (\fli) in the context of their project in the CS2 course.}
  \label{fig:gliToFli}
\end{figure}

Additionally, the fact that they had already been trained to think structurally (i.e., in terms of variables' properties) using \gli helped most of them identify the properties to express and the interval to browse, as they had already been trained to describe zones. Fig.~\ref{fig:gliToFli} illustrates the correlation between students' performance in drawing a \gli and writing the corresponding formal loop invariant. This correlation is statistically significant, and the confidence interval indicates low uncertainty.

{\setlength{\doublerulesep}{0pt}
\setlength{\tabcolsep}{5pt}
\begin{table*}[!h]
  \caption{\centering Translation table from natural language to formal notation, based on the example illustrated in Figure~\ref{fig:findMaxLoopInvariant}.}
  \label{tab:translation}
  \begin{center}
    \resizebox{0.8\textwidth}{!}{%
    \begin{tabular}{l|l}
      \hline
        \textbf{Natural Language Description} & \textbf{Predicate}\\
      \hline\hline\hline
        The elements are not modified. &  $\regle{unmodified}{\forall k \cdot (0 \leq k < \size \Rightarrow \arraya[k] = \arraya_0[k])}$ \\
      \hline
      \var{i} varies from $0$ to \var{N}. & $\regle{lowerbound}{0} \leq \regle{variter}{\counteri} \leq \regle{varsize}{\size}$ \\
      \hline
       \var{index} contains the position of a maximal value. & $\regle{done}{0 \leq \var{\maxindex} < \var{\size}\ \wedge \forall k \cdot (0 \leq k < \counteri \Rightarrow \arraya[\maxindex] \geq \arraya[k])}$ \\
      \hline
    \end{tabular}
    }
  \end{center}
\end{table*}
}

%% file: cafe.tex
\section{\cafe}\label{sec:cafe}

\begin{figure}[!h]
  \begin{center}
    \includegraphics[width=0.8\columnwidth]{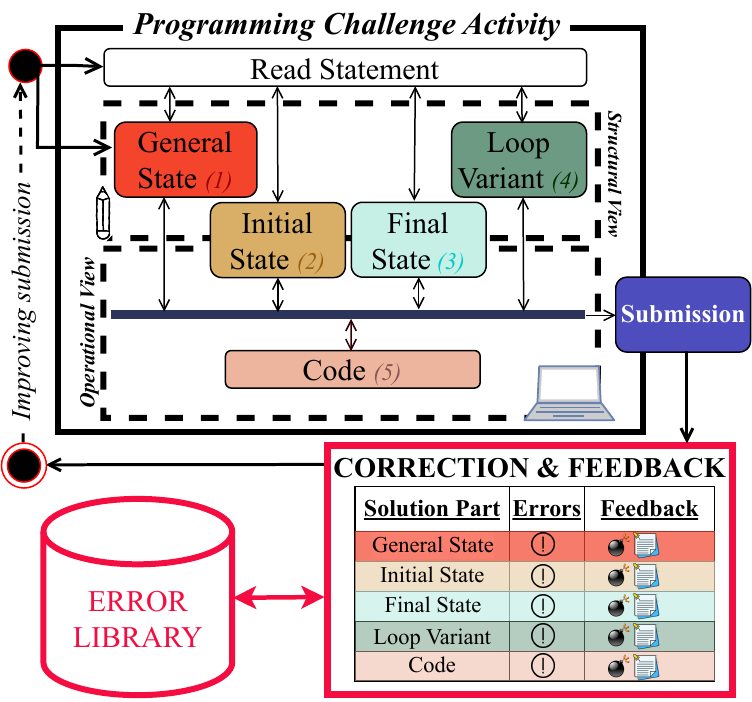}
  \end{center}
  \caption{\centering Activity Diagram showing how students are solving \progchall in \cafe.}
  \label{tools.activityDiagram}
\end{figure}

\begin{figure*}[!h]
  \begin{center}
    \includegraphics[width=14cm]{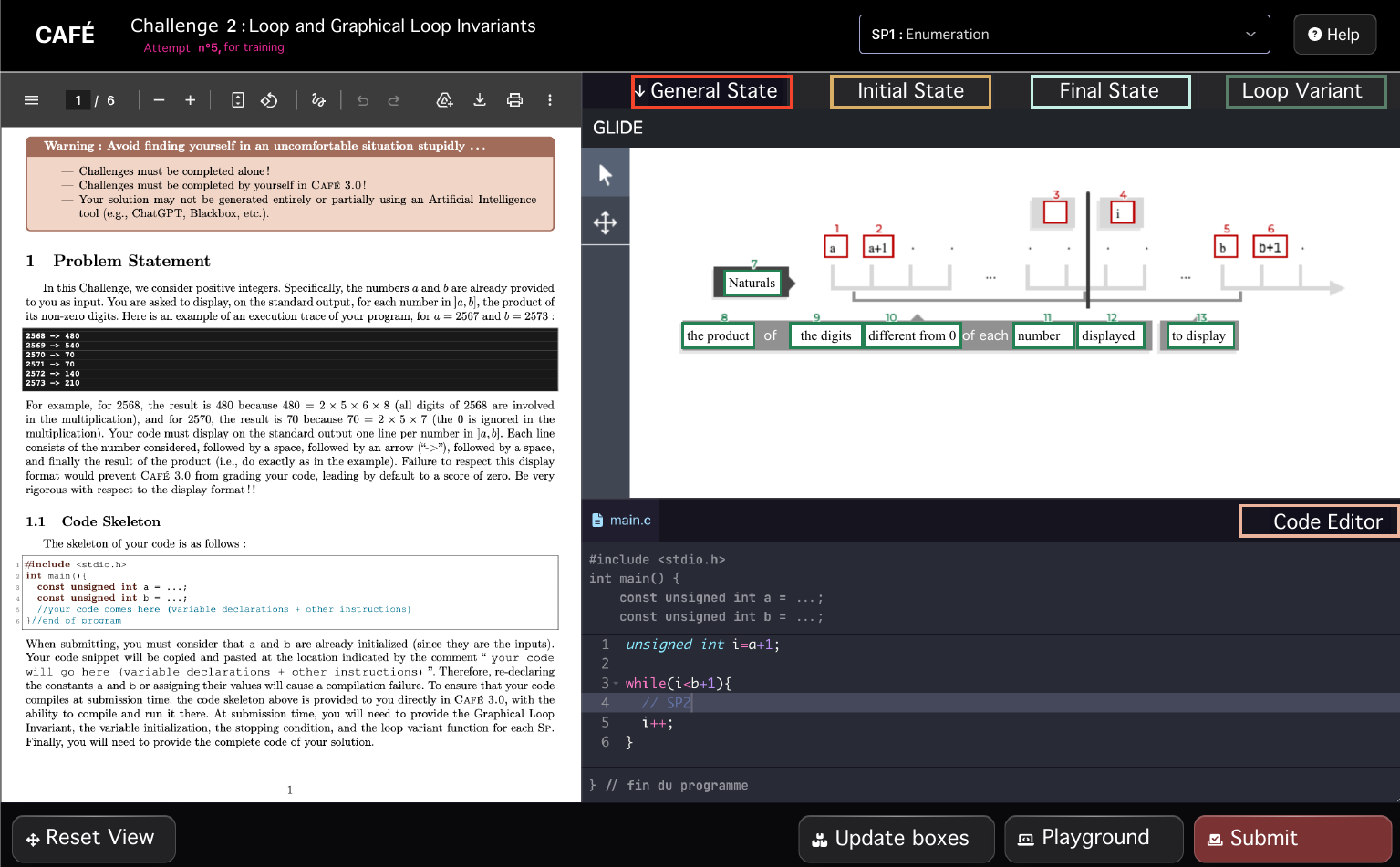}
  \end{center}
  \caption{\centering Screenshot of a \progchall in \cafe.  It also shows how \cafe follows the \glibp methodology with tabs, one for each \solPart.}
  \label{cafe.bgli}
\end{figure*}

\cafe assists students in each subproblem solving by breaking down the expected \glibp solution into specific \solParts.  They are depicted through the upper part of Fig.~\ref{tools.activityDiagram} and, more concretely, through Fig.~\ref{cafe.bgli}, being a screenshot of \cafe. They are aligned with our methodology (see Fig.~\ref{glibp.overview.fig}).  Although students are taught to perform the \solParts from a structural to an operational view, \cafe does not force them to do so.

Focusing on the different \solParts, \cafe also individually outlines them. In Fig.~\ref{cafe.bgli}, the first tab (labeled ``GLI'') contains the \bgli. Each box symbolizes a solution component to be instantiated, like explained in Subsection~\ref{sec:notations}. 
The tab ``Loop Variant'' expects students to provide the loop variant function.  The tabs ``Initial State'' and ``Final State'' allow students to transpose their \gli in specific states to derive the variables' initial value and the loop guard. 
Finally, the code part is represented through a template where students have a placeholder (see ``Code Editor'' on Fig.~\ref{cafe.bgli}). 

%% file: eval.tex
\section{Demo}\label{sec:demo}

\begin{figure*}[!h]
  \begin{center}
    \includegraphics[width=14cm]{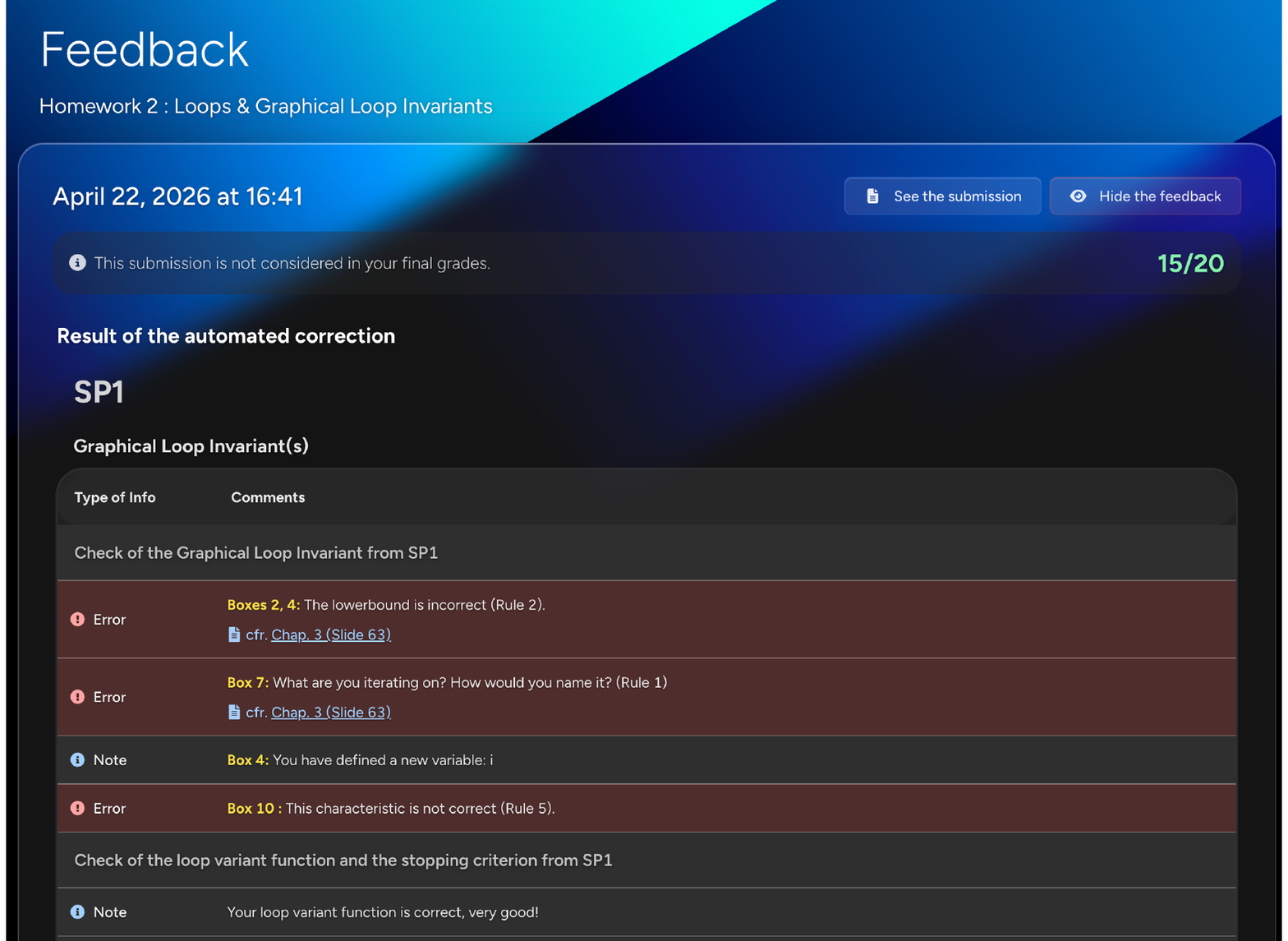}
  \end{center}
  \caption{\centering Screenshot of the feedback page in \cafe.}
  \label{fig:fb}
\end{figure*}

In this demo, we will begin by briefly contextualizing \cafe through an introduction to our \glibp approach (see Subsection~\ref{sec:glibp}), for which we plan to dedicate up to five minutes. We will then explore \cafe from a student's perspective. We will take the first \progchall presented in Table~\ref{tab:challenges_list} as an example, show how to solve the first subproblem (see the interface in Fig.~\ref{cafe.bgli}), and then invite the audience to do the same for the second subproblem. The audience will be able to connect to \cafe (\url{https://cafe.uliege.be/}) as guests. We consider it important that the audience experience \cafe firsthand. 
We will also demonstrate the automated feedback by submitting multiple solutions and discussing the feedback received. In particular, we discuss the trade-off between preserving enough freedom for students to form their own structural view of the solution, and avoiding an overly complex implementation for automating feedback. A screenshot of the feedback page is given in Fig.~\ref{fig:fb}.
In this current version, feedback covers:

\begin{itemize}
  \item The general \gli, based on the content of its boxes (see Fig.~\ref{fig:C1_SP1_1}). Some boxes should contain specific values/variables (typically the lower and upper bounds, often given in the problem statement, or the green boxes composing sentences), while others should respect a relationship with other boxes.
  \item The initial state of the \gli, based on the position of the dividing line(s) and the corresponding initial values specified by the student (see Fig.~\ref{fig:C1_SP1_2}).
  \item The final state of the \gli, based on the position of the dividing line(s) and the corresponding stopping criterion (see Fig.~\ref{fig:C1_SP1_3}).
  \item The loop variant function, based on its general properties (strictly positive, decreasing, etc.) and its consistency with the variables introduced in the \gli and the code.
  \item The code, based on unit tests and consistency with the initial and final state. The consistency check of the loop body is a work in progress.
\end{itemize}

Next, we will switch to the teacher's perspective for about five minutes, during which the audience will discover the encoding platform used to define new \progchalls.

Appendix~\ref{app:allPages} includes a screenshot of each page of \cafe we plan to visit.

For those interested, we may also present our Graphical Loop Invariant Drawing Editor, which allows students to construct \gli diagrams by dragging and dropping components (e.g., the object being iterated over, bars defining zones, text fields, etc.). Additionally, \cafe includes two other modules: the Progress Tracker and the Collaborative Design and Build (CDB) activity. Although not specific to the teaching of FM, both serve as valuable complementary tools. The Progress Tracker provides multiple-choice questions --- covering topics such as \glibp --- on which students receive automated feedback. The CDB module encourages students to think at higher levels of abstraction, rather than focusing solely on writing code.

%% file: extension.tex
\section{Work-in-progress to bridge \gliFull to Formal Loop Invariant}\label{extension}

To bring further automated feedback on \glibpFull, we have implemented a proof-of-concept prototype where the \gli and corresponding loop implementation are translated into Dafny for code verification. This involves translating \gli{}s into Dafny predicates by extracting three core components: ($i$) the object(s) being iterated over such as arrays, ($ii$) the zone delimiters, including bounds and dividing lines, and ($iii$) the textual descriptions of the zones, including references to accumulator variables. We map sentences to predicates using a predefined library of predicates characterized by corresponding sentences. And we integrate the resulting annotations into the Dafny instructions generated from the original C code. We have already tested this approach on ten representative CS1 problems, for instance, finding a maximum value in an array.

%% file: conclusion.tex

\section{Conclusion}\label{sec:concl}

We have presented \cafe, a learning platform that supports the teaching of \glibpFull in CS1. By combining a structured visual representation of the loop invariant with personalized automated feedback, \cafe helps students develop structural thinking skills early in their education, laying the foundations for Formal Methods. The platform has been deployed in our CS1 course over multiple academic years, with students completing three graded \progchalls per semester.

Looking ahead, we are working on a proof-of-concept prototype that automatically translates a \gli and its corresponding C implementation into Dafny, enabling formal verification and richer feedback. We see this as a promising step towards closing the gap between informal diagrammatic reasoning and rigorous formal methods. We welcome collaboration with other institutions interested in adopting or extending \cafe for their own FM teaching contexts.

%% file: appendix.tex
\section{All the pages from \cafe}\label{app:allPages}

This appendix presents screenshots of each page composing the demo walkthrough in \cafe. The sequence follows the typical workflow a student encounters when solving \progchall: from the main entry point of the platform, through the successive steps of constructing a \gli and deriving the corresponding code.

\begin{figure*}[!h]
  \begin{center}
    \includegraphics[width=14cm]{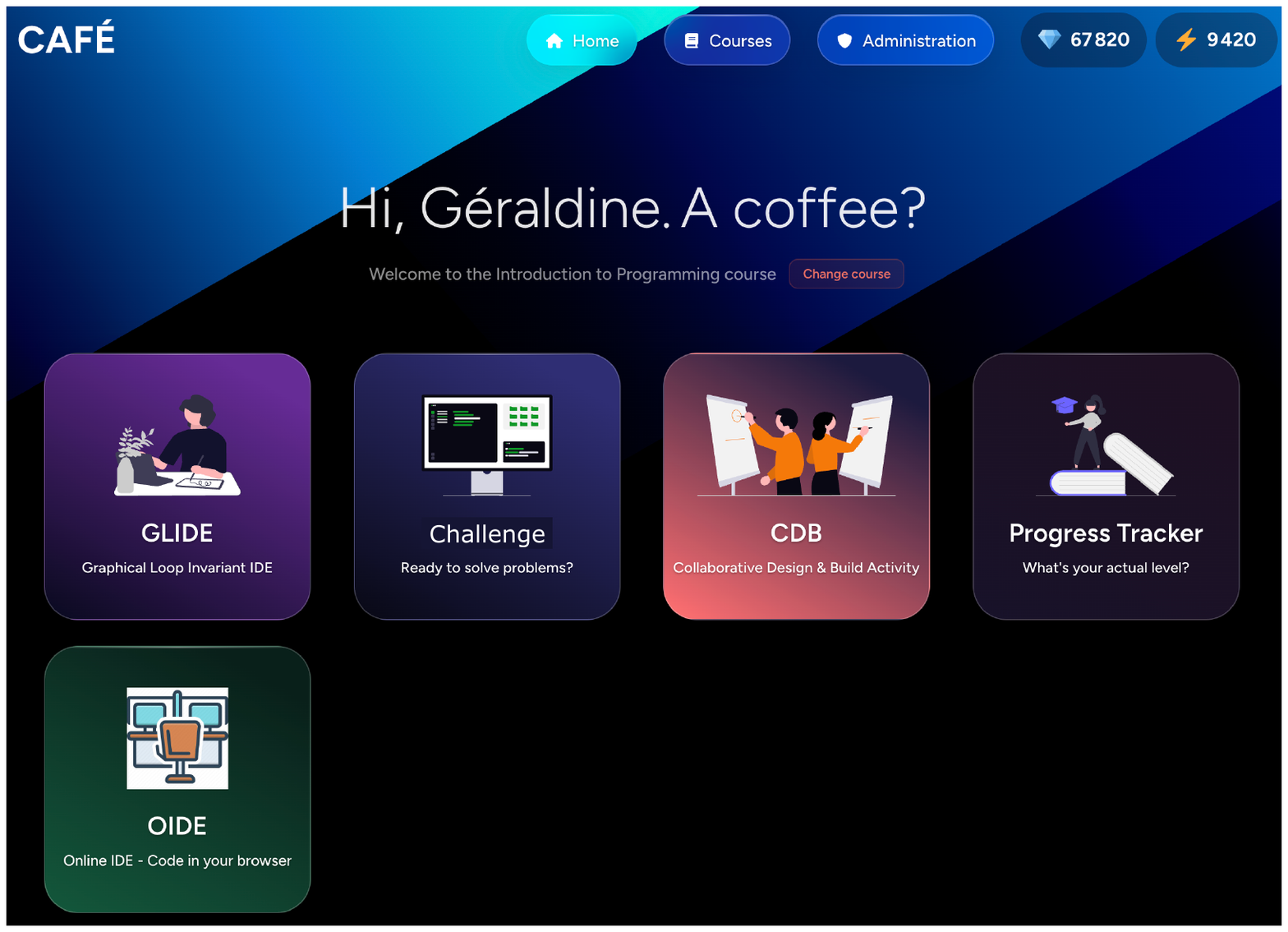}
  \end{center}
  \caption{\centering Main page in \cafe.}
  \label{fig:mainPage}
\end{figure*}

Upon logging in, students access the main page (Fig.~\ref{fig:mainPage}). From there, they can visit five modules. The one of interest for this demo is ‘‘Challenge''.

\begin{figure*}[!h]
  \begin{center}
    \includegraphics[width=14cm]{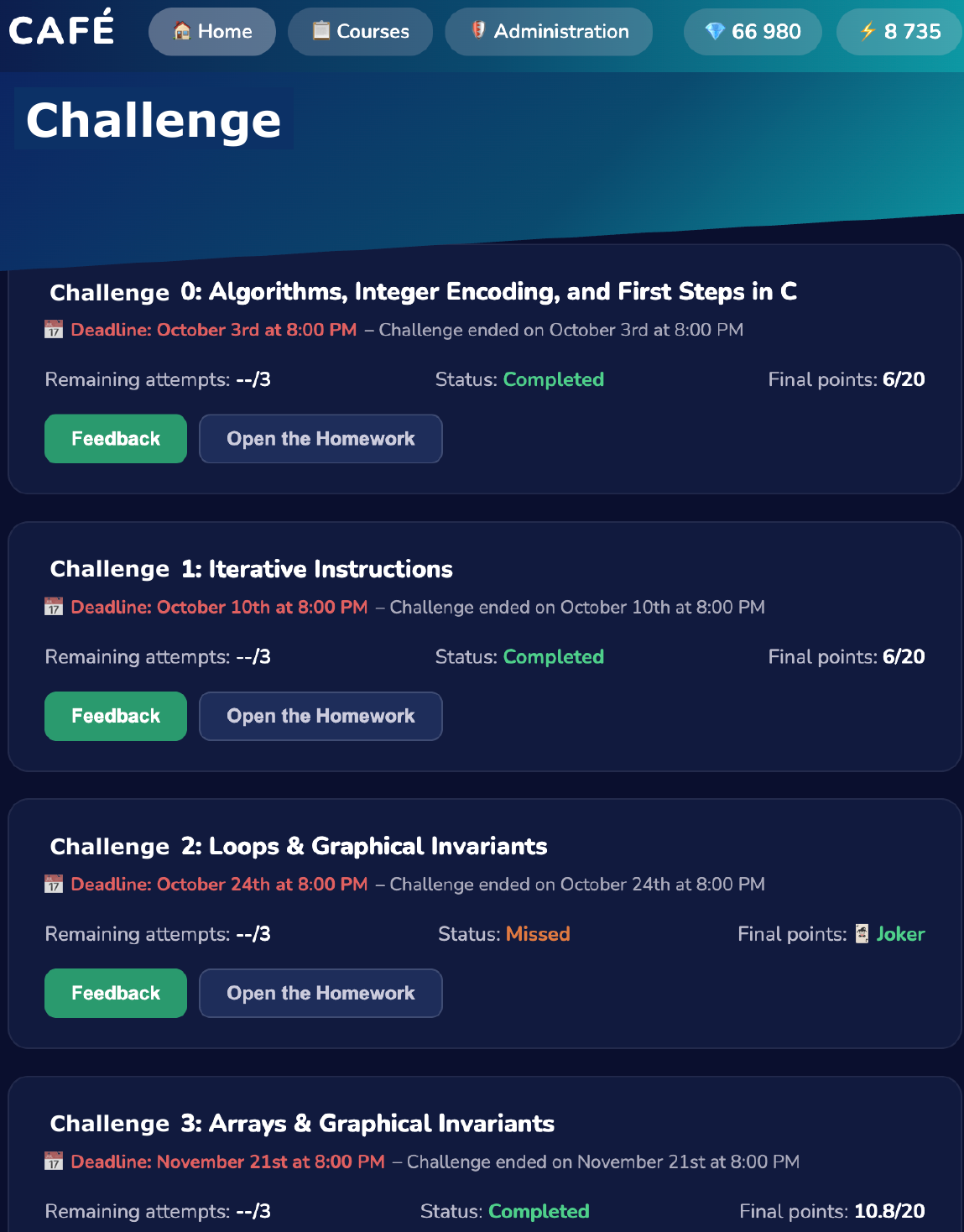}
  \end{center}
  \caption{\centering Page listing the \progchalls available in \cafe.}
  \label{fig:listChals}
\end{figure*}

After clicking on ‘‘Challenge'', students access the page listing all the available \progchalls (see Fig.~\ref{fig:listChals}). Each entry provides a brief description of the problem, allowing students to select the one they wish to attempt. In our demo, we focus on \progchall 2 (being the first challenge mobilizing \glibp).

\begin{figure*}[!h]
  \begin{center}
    \includegraphics[width=14cm]{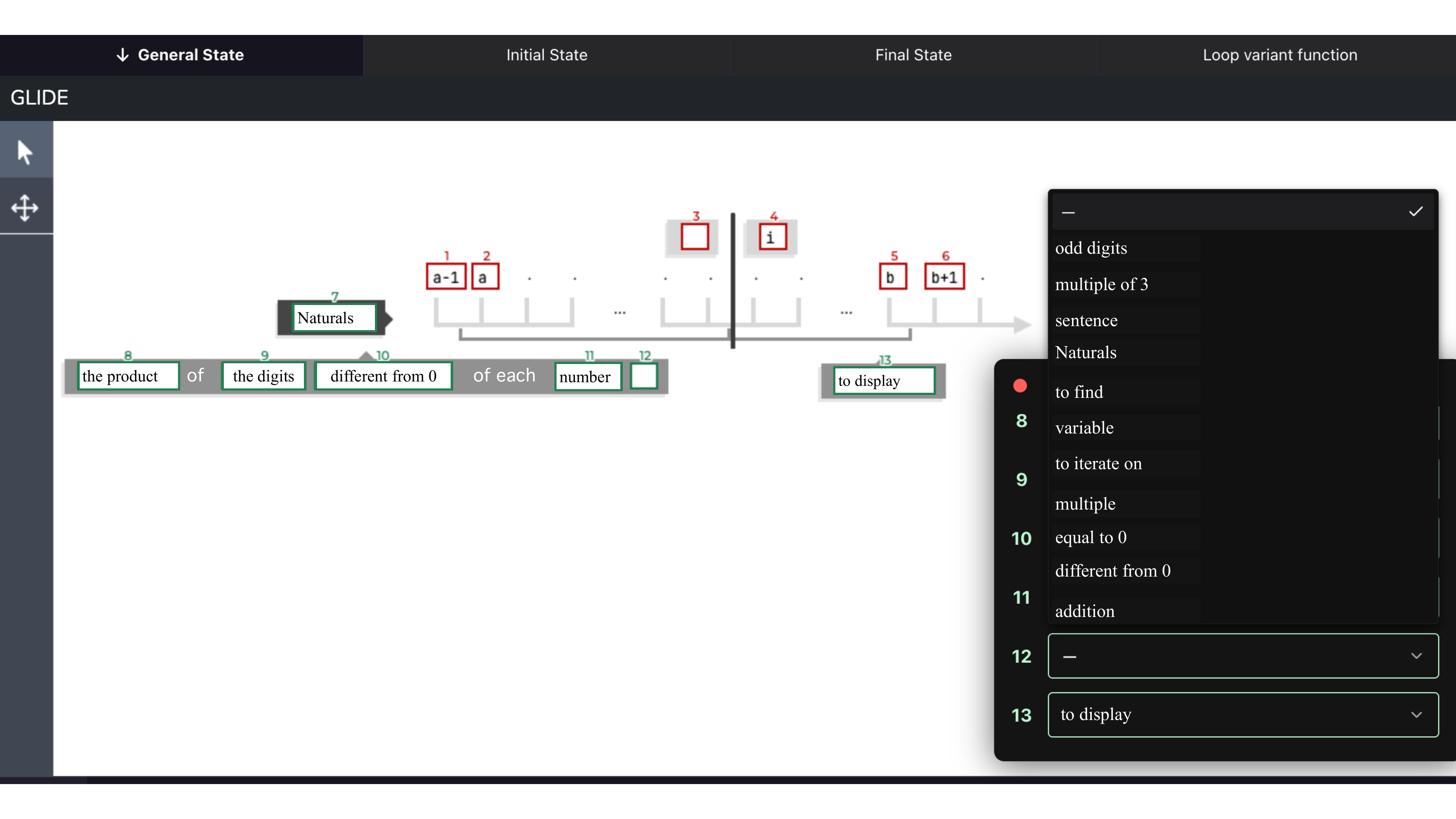}
  \end{center}
  \caption{\centering Page for completing the \bgli for the first subproblem of \progchall 1 in \cafe.}
  \label{fig:C1_SP1_1}
\end{figure*}

The first step of the solving process (Fig.~\ref{fig:C1_SP1_1}) asks students to fill in the \bgli corresponding to the general state of the loop. This is the central artefact of the \glibp approach, from which all subsequent steps are derived.

\begin{figure*}[!h]
  \begin{center}
    \includegraphics[width=14cm]{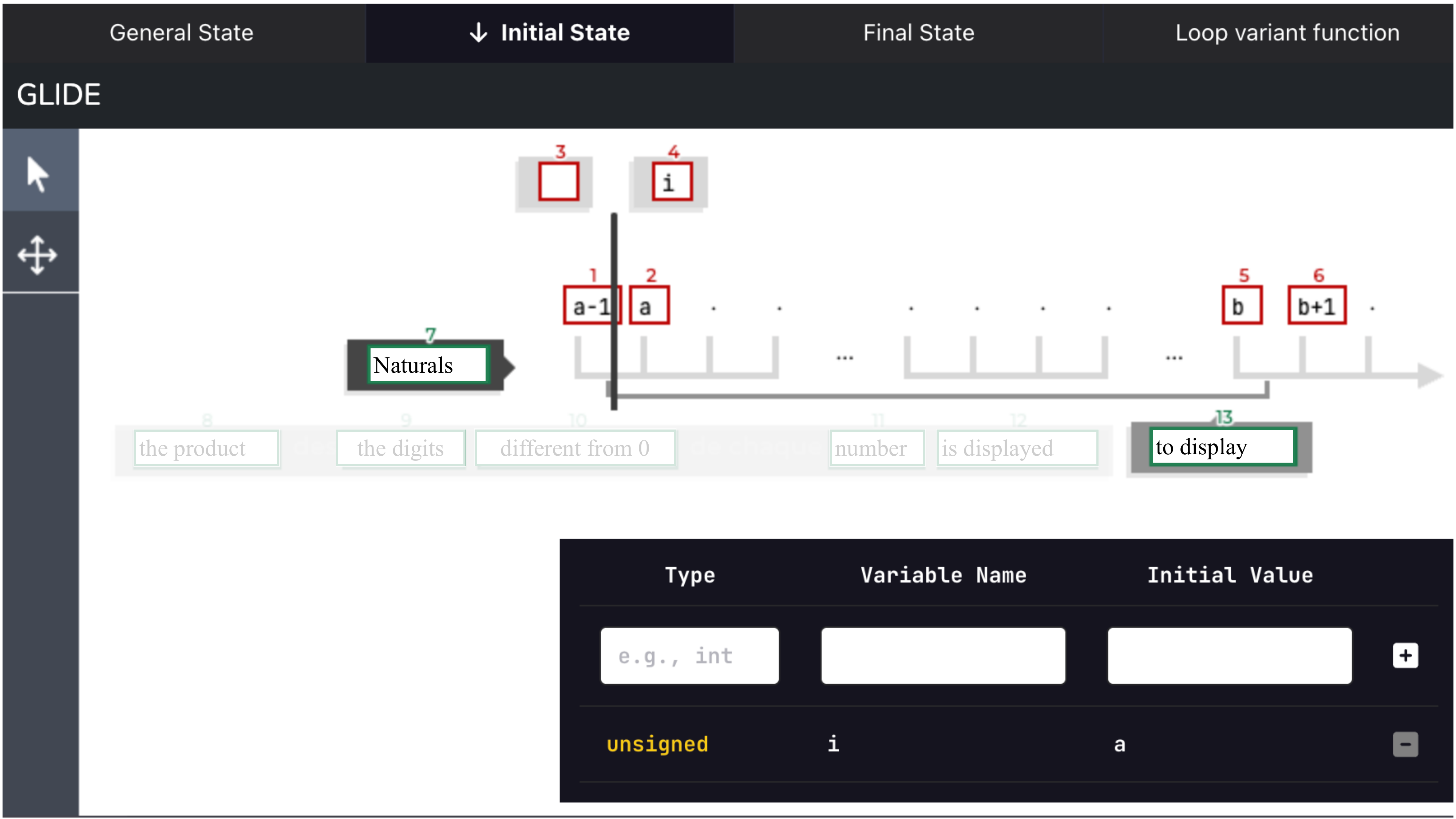}
  \end{center}
  \caption{\centering Page for specifying the \initState for the first subproblem of \progchall 1 in \cafe.}
  \label{fig:C1_SP1_2}
\end{figure*}

Once the general \gli is established, students instantiate it into the \initState (Fig.~\ref{fig:C1_SP1_2}), capturing the state of the variables before the loop begins. This step directly informs the variable declaration and initializations in the code, that students should in the lower-right frame.

\begin{figure*}[!h]
  \begin{center}
    \includegraphics[width=14cm]{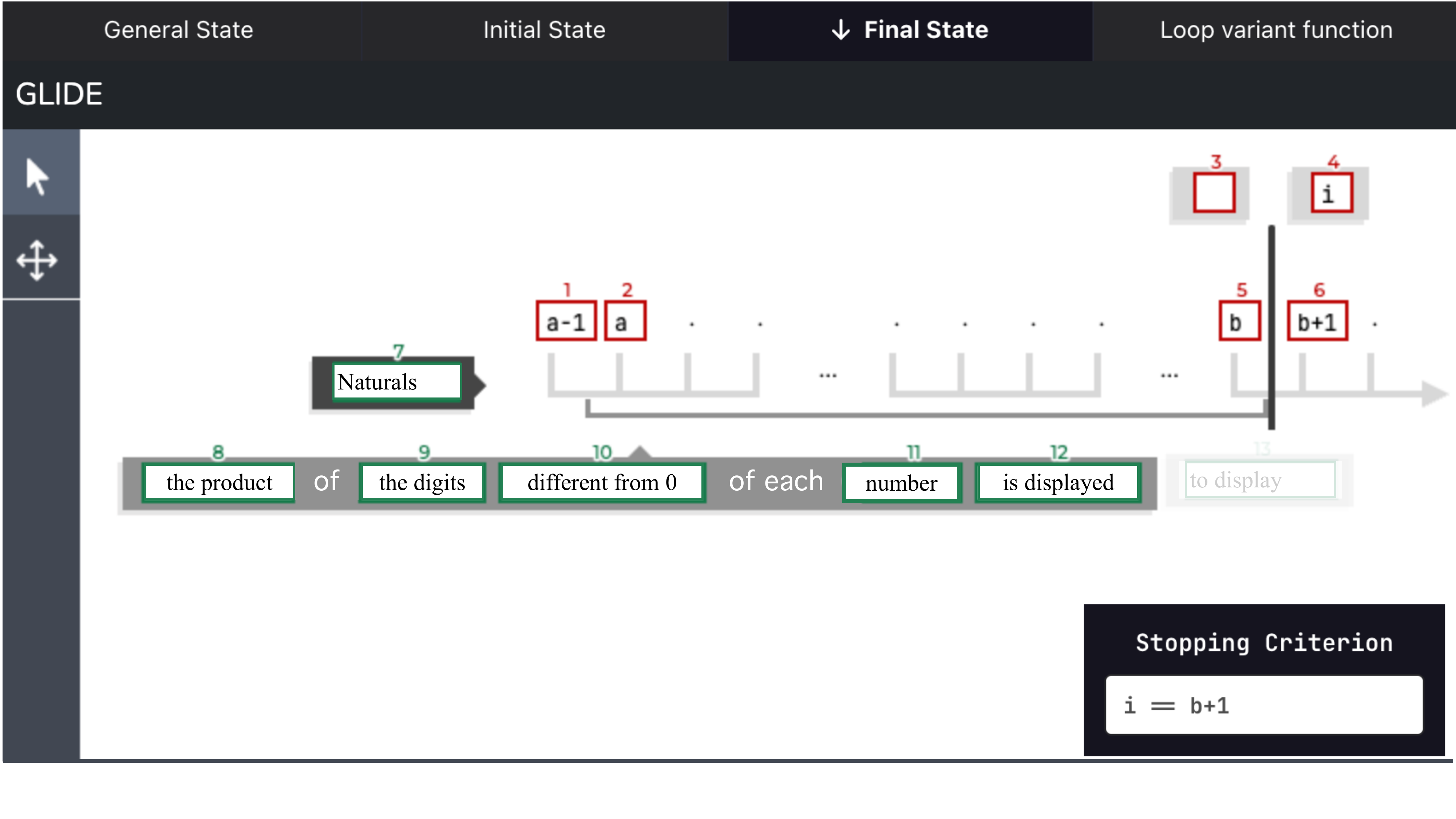}
  \end{center}
  \caption{\centering Page for specifying the \finalState for the first subproblem of \progchall 1 in \cafe.}
  \label{fig:C1_SP1_3}
\end{figure*}

Students then instantiate the \gli into the \finalState (Fig.~\ref{fig:C1_SP1_3}), representing the state of the variables upon loop exit. From this, they derive the loop's \critere they should write in the lower-right frame.

\begin{figure*}[!h]
  \begin{center}
    \includegraphics[width=14cm]{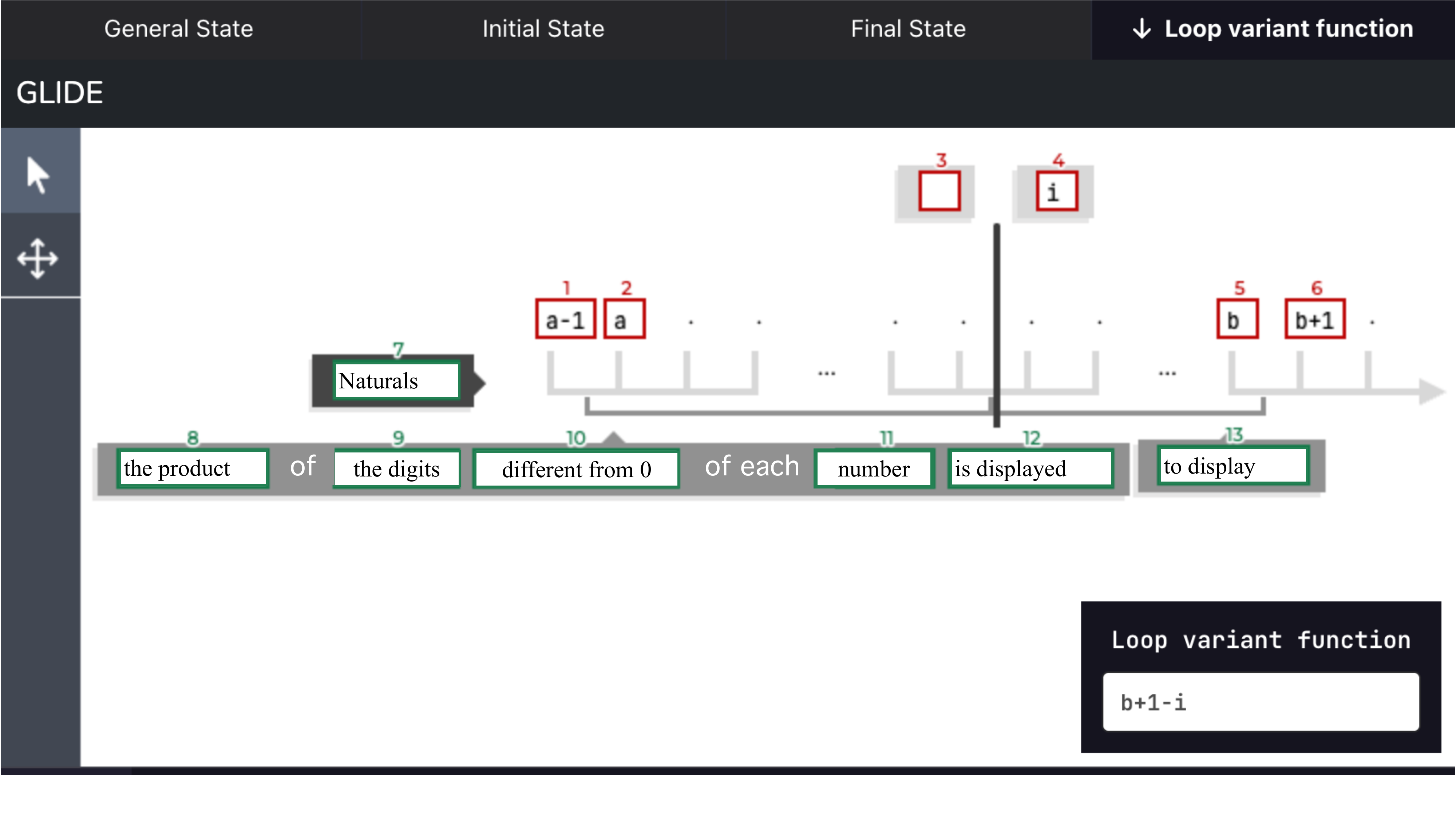}
  \end{center}
  \caption{\centering Page for determining the loop variant function for the first subproblem of \progchall 1 in \cafe.}
  \label{fig:C1_SP1_4}
\end{figure*}

Fig.~\ref{fig:C1_SP1_4} shows the page dedicated to the loop variant function. Students must provide an expression that strictly decreases at each iteration and is bounded below, thereby guaranteeing that the loop terminates. It corresponds to the size of the green zone that should be reduced at each iteration.

\begin{figure*}[!h]
  \begin{center}
    \includegraphics[width=14cm]{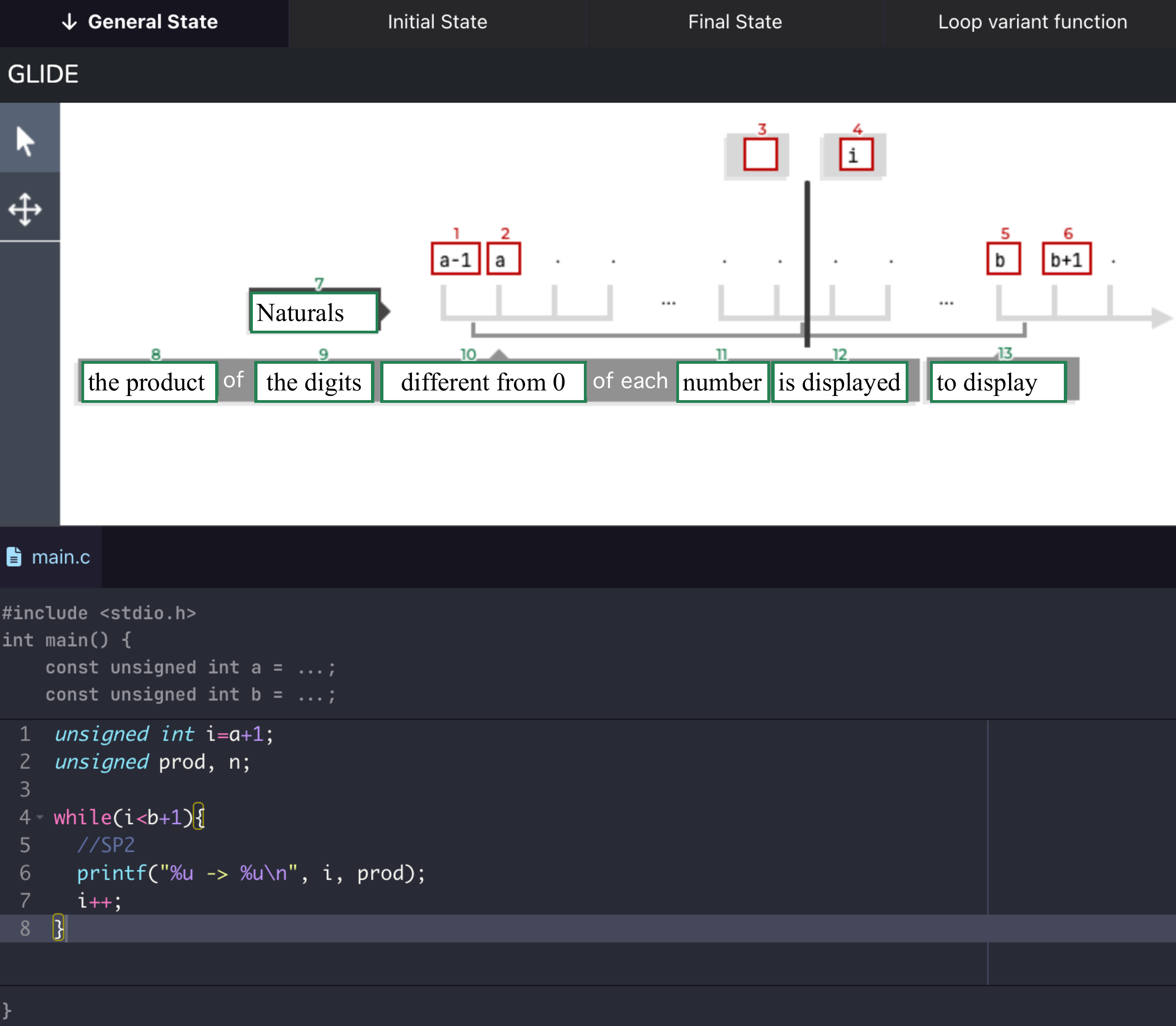}
  \end{center}
  \caption{\centering Page for writing the code for the first subproblem of \progchall 1 in \cafe.}
  \label{fig:C1_SP1_5}
\end{figure*}

Finally, Fig.~\ref{fig:C1_SP1_5} shows the code submission page, where students write the loop implementation derived from their \gli. \cafe evaluates the submitted code and provides personalized automated feedback, allowing students to iteratively refine both their \gli and their code. The feedback page is shown in Fig.~\ref{fig:fb}.

%% file: Bibliography.bib
@string{FMTEA = "Proc. Formal Method Teaching Workshop ({FMTea})"}

@string{EDUCON = "Proc. {IEEE} Global Engineering Education Conference ({EDUCON})"}

@string{INFEDU = "Informatics in Education"}

@string{SIGCSE = "Proc. {ACM} Technical Symposium on Computer Science Education ({SIGCSE})"}

@string{ITiCSE = "Proc. {ACM} Conference on Innovation and Technology in Computer Science Education ({ITiCSE})"}

@article{li_otherApproachWithRec,
author = "Morazán, M.",
year = "2020",
month = "August",
pages = "1--18",
title = "How to Design While Loops",
volume = "321",
number = "13",
journal = "Electronic Proceedings in Theoretical Computer Science",
doi = "10.4204/EPTCS.321.1",
}

@article{3_FormalThinking,
author = "Dongol, B. and Dubois, C. and Hallerstede, S. and Hehner, E. and Morgan, C. and M{\"u}ller, P. and Ribeiro, L. and Silva, A. and Smith, G. and de Vink, E.",
title = "On Formal Methods Thinking in Computer Science Education",
year = "2024",
doi = "10.1145/3670419",
journal = "Format Aspects of Computing",
month = "May",
volume = "37",
number = "1",
pages = "1--23",
}

@InProceedings{moreFM,
author="Kamburjan, E. and Gr{\"a}tz, L.",
title="Increasing Engagement with Interactive Visualization: Formal Methods as Serious Games",
booktitle=FMTEA,
year="2021",
month = "November",
}

@InProceedings{moreFMTea,
author="Zhumagambetov, R.",
title="Teaching Formal Methods in Academia: A Systematic Literature Review",
booktitle="Proc. Formal Methods -- Fun for Everybody",
year="2021",
month = "December",
}

@article{FMimportant,
author = "Broy, M. and Brucker, A.~D. and Fantechi, A. and Gleirscher, M. and Havelund, K. and Kuppe, M.~A. and Mendes, A. and Platzer, A. and Ringert, J.~O. and Sullivan, A.",
title = "Does Every Computer Scientist Need to Know Formal Methods?",
year = "2024",
volume = "37",
number = 1,
journal = "Format Aspect of Computing",
month = "December",
pages = "1--17",
}

@BOOK{morgan_book,
title = "Programming from Specifications",
author = "Morgan, C.",
year = "1990",
publisher = "Prentice-Hall",
}

@BOOK{gries_book,
title = "The Science of Programming",
author = "Gries, D.",
publisher = "Springer",
year = "1987",
}

@ARTICLE{hoare,
author = "Hoare, C. A. R.",
title = "An Axiomatic Basis for Computer Programming",
journal = "Communications of the {ACM}",
volume = "12",
number = "10",
pages = "576--580",
year = "1969",
month = "October",
}

@book{dijkstra,
title = "A Discipline of Programming",
author = "Dijkstra, E. W.",
year = "1976",
publisher = "Prentice-Hall, Inc.",
}

@INPROCEEDINGS{invariants_pictures,
title = "Pictures as Invariants",
author = "Astrachan, O.",
booktitle = SIGCSE,
year = "1991",
month = "March",
doi = "10.1145/107004.107026",
}

@INPROCEEDINGS{teaching_loop_invariant,
title = "Teaching Loop Invariants to Beginners by Examples",
author = "Tam, W. C",
booktitle = SIGCSE,
year = "1992",
month = "March",
doi = "10.1145/135250.134530",
}

@Book{Anderson2002,
editor    = "Anderson, M. and Meyer, B. and Olivier, P.",
publisher = "Springer London",
title     = "Diagrammatic Representation and Reasoning",
year      = "2002",
doi       = "10.1007/978-1-4471-0109-3",
}

@ARTICLE{invariant_edu,
author = "Mannila, L.",
title = "Invariant Based Programming in Education -- An Analysis of Student Difficulties",
journal = INFEDU,
volume = "9",
number = "1",
pages = "115--132",
year = "2010",
month = "April",
doi = "10.15388/infedu.2010.07",
}

@inproceedings{cafe2,
author = "Brieven, G. and Malcev, L. and Donnet, B.",
title = "Practicing Abstraction Skills Through Diagrammatic Reasoning Over {CAF\'E} 2.0",
booktitle = "Proc. {IEEE} Global Engineering Education Conference ({EDUCON})",
year = "2024",
}

@inproceedings{4_glibp,
author = "Brieven, G. and Li\'enardy, S. and Malcev, L. and Donnet, B.",
booktitle = FMTea,
month = "March",
title = "Graphical Loop Invariant Based Programming",
year = "2023",
}

@article{structuralView,
author = "Mirolo, C. and Izu, C. and Lonati, V. and Scapin, E.",
title = "Abstraction in Computer Science Education: An Overview",
journal = INFEDU,
volume = "20",
number = "4",
year = "2022",
pages = "615--639",
doi = "10.15388/infedu.2021.27",
month = "December"
}

@inproceedings{loop_invariant_misconceptions,
author = "Fowler, M. and Kraemer, E. and Sitaraman, M. and Hollingsworth, J. E.",
title = "Tool-Aided Loop Invariant Development: Insights into Student Conceptions and Difficulties",
doi = "10.1145/3430665.3456351",
booktitle = ITiCSE,
year = "2021",
month = "June/July",
}

@inproceedings{tool_proofs,
author = "Knobelsdorf, M. and Frede, C. and Bohne, S. and Kreitz, C.",
title = "Theorem Provers as a Learning Tool in Theory of Computation",
year = "2017",
isbn = "9781450349680",
publisher = "Association for Computing Machinery",
address = "New York, NY, USA",
doi = "10.1145/3105726.3106184",
booktitle = "Proceedings of the 2017 ACM Conference on International Computing Education Research",
pages = "83–92",
numpages = "10",
series = "ICER '17"
}
